\documentclass[traditabstract]{aa} 
\usepackage{graphicx}
\usepackage{supertabular}
\usepackage{txfonts}
\begin{document}

\title{Improved prospects for the detection\\ of new Large Magellanic Cloud planetary nebulae\thanks{Based on observations made with the Wide Field Imager of the Max-Planck-ESO 2.2m telescope at La Silla Observatory under program ID 076.C-0888, processed and released by the ESO VOS/ADP group, VISTA at Paranal Observatory under program ID 179.B-2003 and the Wide Field Spectrograph of the Australian National University 2.3m telescope.}}

   \author{Brent Miszalski
           \inst{1}
           \and
           Ralf Napiwotzki
           \inst{1}
           \and
           Maria-Rosa L. Cioni
           \inst{1,2}\thanks{Research Fellow of the Alexander von Humboldt Foundation}
           \and
           Jundan Nie
           \inst{3,4}
          }

\institute{Centre for Astrophysics Research, STRI, University of Hertfordshire, College Lane Campus, Hatfield AL10 9AB, UK\\
\email{b.miszalski@herts.ac.uk}
         \and
         University Observatory Munich, Scheinerstrasse 1, D-81679, M\"unchen, Germany
         \and
         Research School of Astronomy and Astrophysics, Australian National University, Cotter Road, Weston Creek ACT 2611, Australia
         \and
         Department of Astronomy, Beijing Normal University, Beijing, 100875, China
         }
   \date{Received -; accepted -}

\abstract{The Large Magellanic Cloud (LMC) contains the nearest large extragalactic population of planetary nebulae (PNe). A shallow viewing angle and low interstellar reddening towards the LMC potentially means a larger, more complete flux-limited population can be assembled than for any other galaxy. These advantages appear to be reflected by the small gap between the catalogued ($\sim$700 PNe) and estimated ($1000\pm250$ PNe) population size. With more detailed multi-wavelength studies the catalogued number of LMC PNe may fall, potentially widening this gap. We demonstrate here that the gap can be further bridged with improved optical and near-infrared imaging surveys. We present three [O~III]-selected PNe discovered from ESO WFI observations of the 30 Doradus region and one serendipitous discovery from near-infrared Vista Magellanic Cloud (VMC) survey observations. The WFI PNe have resolved [O~III] and H$\alpha$ nebulae that verify their PN nature and their [O~III] fluxes place them 6--7 mag ($m_{5007}=20$--21 mag) fainter than the bright-end of the planetary nebula luminosity function (PNLF). Their faintness, small angular size and surrounding complex emission-line background explains why previous H$\alpha$ surveys of the region did not select them. We estimate there may be as many as 50--75 similar PNe awaiting discovery in the central $5\times5$ degrees of the LMC. 
The VMC survey routinely detects PNe as red resolved nebulae that may allow some of this expected population to be recovered without traditional narrow-band imaging surveys. We demonstrate this potential with the first new VMC-selected PN which has a rare Wolf-Rayet [WC9]--[WC11] central star. 

   }
   \maketitle
   \section{Introduction}
   Planetary nebulae (PNe) are a brief ($\sim$10$^4$ yr) phase in the evolution of low-intermediate mass stars after the asymptotic giant branch (AGB). A remnant pre-white dwarf core ionises the ejected gaseous envelope producing a strong emission line spectrum enabling PNe to be visible at even extragalactic distances. The Magellanic Clouds are a unique laboratory for the study of PNe where the favourable inclination (van der Marel \& Cioni 2001) and low interstellar reddening allows for a large and highly complete PN population to be assembled. This cannot be achieved in the heavily reddened Milky Way or in more distant members of the Local Group where current observations are only sensitive to brighter PNe. Magellanic PNe (MCPNe) can be studied in detail thanks to their close proximity (e.g. Shaw et al. 2006) and the known distance means fundamental parameters can be measured in astrophysically useful units (e.g. nebula radii in pc). MCPNe are also powerful probes of low-metallicity AGB nucleosynthesis, the rotation curve as kinematic tracers and spatial variations of chemical abundances (see e.g. Shaw 2006; Stanghellini 2009). The standard candle [O~III] planetary nebula luminosity function (PNLF) was also formulated from MCPNe (Jacoby 1980, 1989; Ciardullo et al. 1989) and MCPNe remain an important stepping stone towards understanding why the PNLF works in more distant galaxies (Ciardullo 2010).

   All these applications depend upon the discovery of the maximum number of PNe possible to measure generally weak trends or features across the Magellanic Clouds. Knowing the actual number of PNe is also essential to check the uncertain theoretical estimates of the total MCPNe population. Jacoby (1980) and Peimbert (1990) estimated the total LMC and SMC population size to be $1000\pm250$ and $290\pm80$ PNe, respectively. Jacoby \& De Marco (2002) more recently estimated $\sim$1140 and $\sim$216, respectively, based on a deep census of SMC PNe. Our interest here is in the much larger LMC population for which there are currently catalogued $\sim$280 PNe from multiple surveys (Leisy et al. 1997) and $\sim$450 PNe from Reid \& Parker (2006a, 2006b; hereafter RP2006a and RP2006b). The discoveries made by RP2006b go a long way towards reaching the expected $\sim$$1000\pm250$ total, though it is possible that many or even all of the 169 PNe RP2006b classified as `likely' or `possible' may not turn out to be PNe with further study. Invariably H$\alpha$ selected samples of PNe contain a large amount of non-PNe that must be removed based on detailed multi-wavelength criteria (Frew \& Parker 2010). Reid \& Parker (2010, hereafter RP2010) revised their classifications of 26 candidates as part of their PNLF study and potentially even more may yet be removed.

   The VISTA Magellanic Cloud (VMC) survey\footnote{http://star.herts.ac.uk/$\sim$mcioni/vmc/} is obtaining deep $YJK_s$ photometry of the Magellanic Clouds and Bridge at sub-arcsecond resolution (Cioni et al. 2011). VMC observations will be a powerful tool to identify non-PNe amongst all catalogued MCPNe. The first year of observations include six LMC tiles containing 98 objects catalogued as true, likely or possible PNe. Miszalski et al. (2011) found at most only 48/98 (49\%) were bona-fide PNe after an assessment of all available multi-wavelength and time-series observations. These figures are heavily biased by the 30 Doradus region and its associated complex emission-line background which complicates the identification of true emission-line objects like PNe. The VMC survey will also allow new PNe to be found that will contribute to the overall tally of MCPNe.

   This paper describes a systematic search for new PNe in a 63$\times$63 arcmin$^2$ region near 30 Doradus with narrow-band optical imaging data, as well as the serendipitous discovery of a new PN identified solely from the VMC survey data. Section \ref{sec:method} describes the available data and our applied search methods. Section \ref{sec:results} presents three new [O~III]-selected PNe and one new VMC-selected PN and their basic properties. We conclude in Sect. \ref{sec:end}. 

\section{Data and search method}
\label{sec:method}
   \subsection{A systematic search of optical narrow-band imaging}
   To conduct our search for new PNe we utilise the publicly-available $B$, $V$, [O~III] and H$\alpha$ images taken under non-photometric conditions with the Wide Field Imager (WFI) of the ESO 2.2-m telescope under program ID 076.C-0888.\footnote{http://archive.eso.org/archive/adp/ADP/30\_Doradus} The data were reduced by ESO following Nonino et al. (1999) and cover a large $63\times63$ arcmin$^2$ region centred near 30 Doradus ($\alpha_\mathrm{J2000}=05^\mathrm{h}37^\mathrm{m}54.7^\mathrm{s}$, $\delta_\mathrm{J2000}=-69^\circ21'55''$). This region is made upo of four separate sub-fields: 30Dor1, 30Dor2, 30Dor3 and 30Dor4, all sampled at 0.238\arcsec/pixel. The world coordinate system (WCS) solution is accurate to 0.2\arcsec\ (rms) over the entire area. Table \ref{tab:wfi} reproduces the coordinates and observation dates of the exposures taken, their total exposure times in each filter and the average measured stellar full-widths at half maximum (FWHM) as recorded at the data products webpage. The central wavelengths and FWHMs of the filters were 451.1/133.5 nm (B/123), 539.6/89.4 nm (V/89) 502.4/8.0 nm (OIII/8) and 658.8/7.4 nm (Halpha/7). 

   \begin{table*}
      \centering
      \caption{Log of ESO WFI observations made under program ID 076.C-0888.}
      \label{tab:wfi}
      \begin{tabular}{llllllll}
        \hline\hline
        Field	& Filter &	RA & Dec. &	Total Exposure Time & Exposures & FWHM & Observed\\
               &        &  (J2000) & (J2000) & (seconds) & & (\arcsec) & (YYYY-MM-DD)\\ 
         \hline
30Dor1 & B/123 & 05 40 49.0 & $-$69 07 01.7 & 1500 & 5 & 0.95 & 2006-01-29\\
30Dor1 & OIII/8 & 05 40 49.3 & $-$69 06 49.2 & 1500 & 5 & 0.82 & 2006-01-29\\
30Dor1 & V/89 & 05 40 49.4 & $-$69 07 23.7 & 1500 & 5 & 0.83 & 2006-01-28\\
30Dor1 & Halpha/7 & 05 40 48.5 & $-$69 06 47.3 & 4800 & 4 & 0.73 & 2006-01-27\\
30Dor2 & B/123 & 05 35 11.2 & $-$69 06 45.4 & 1500 & 5 & 0.92 & 2006-01-29\\
30Dor2 & OIII/8 & 05 35 12.2 & $-$69 06 32.1 & 1500 & 5 & 0.85 & 2006-01-29\\
30Dor2 & V/89 & 05 35 12.0 & $-$69 07 11.6 & 1500 & 5 & 0.78 & 2006-01-28\\
30Dor2 & Halpha/7 & 05 35 11.8 & $-$69 06 31.2 & 6000 & 5 & 0.85 & 2006-01-27\\
30Dor3 & B/123 & 05 40 49.8 & $-$69 37 03.5 & 1500 & 5 & 1.00 & 2006-01-29\\
30Dor3 & OIII/8 & 05 40 46.6 & $-$69 36 57.0 & 1200 & 4 & 0.90 & 2006-01-29\\
30Dor3 & V/89 & 05 40 49.7 & $-$69 37 30.4 & 1500 & 5 & 0.75 & 2006-01-28\\
30Dor3 & Halpha/7 & 05 40 49.2 & $-$69 37 01.6 & 6000 & 5 & 1.17 & 2006-01-27\\
30Dor4 & B/123 & 05 35 04.2 & $-$69 36 39.1 & 1500 & 5 & 0.88 & 2006-01-29\\
30Dor4 & OIII/8 & 05 35 04.9 & $-$69 36 28.0 & 1500 & 5 & 1.12 & 2006-01-29\\
30Dor4 & V/89 & 05 35 04.4 & $-$69 37 15.7 & 1800 & 6 & 0.85 & 2006-01-28\\
30Dor4 & Halpha/7 & 05 35 02.8 & $-$69 37 27.4 & 4800 & 4 & 0.84 & 2006-01-28\\
\hline
      \end{tabular}
   \end{table*}

Our strategy was to thoroughly search the WFI data in the form of a three-colour image of the four sub-fields made from H$\alpha$ (red), [O~III] (green) and $B$ (blue) images. In this combination bona-fide PNe typically have a telltale yellow--green hue from their strong H$\alpha$ and [O~III] emission, but there remains some sensitivity to H$\alpha$ emitting point-sources and nebulae if there is no [O~III] emission. The very large size of the four sub-fields ($\sim$$8500\times8500$ pixels) called for a methodical approach when visualising the data to ensure no parts were left unexamined. This was achieved using a custom-made plugin developed by one of us (BM) for the \textsc{ds9} program (Joye \& Mandel 2003). When an image is loaded the plugin divides it up into manageable sub-frames that fit within the \textsc{ds9} graphical window dimensions. A 10\% overlap between sub-frames ensured no parts of the images were missed when navigating from sub-frame to sub-frame. 

While browsing the images we made use of the many different scaling modes provided by \textsc{ds9} to separately target faint emission at high contrast and point-source or compact emitters at low contrast that diminishes the high nebular background. The \textsc{ds9} region functionality was used to overlay previously catalogued PNe and to record the position of new candidates. A total of 297 candidates were initially selected based on suspected [O~III] and/or H$\alpha$ emission. Candidates were then visualised in a web page that included the aforementioned WFI colour-composite image and a VMC colour-composite image made from $K_s$ (red), $J$ (green) and $Y$ (blue). Only three promising PN candidates were found with almost all the rest turning out to be AGB stars (e.g. Miras) whose TiO bands give the false impression of H$\alpha$ emission. These were easily discerned as very bright sources in the VMC colour-composite images ($K_s\la14$ mag). We also recovered the nebula of Supernova 1987A and the [O~III]-bright pulsar wind nebula of PSR B0540$-$69.3. A few peculiar nebulae around luminous stars were also recovered (e.g. Henize 1956; Weis et al. 1997). As our focus is solely on the detection of new PNe we refrain from further discussion of these objects.

\subsection{A serendipitous VMC discovery}
\begin{figure}
   \begin{center}
   \includegraphics[scale=0.288]{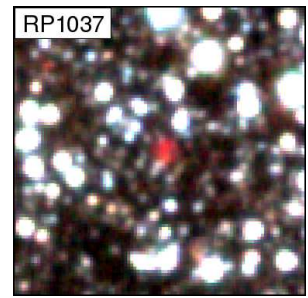}
   \end{center}
   \caption{VMC colour-composite image of the PN RP1037 (RP2006b) made from stacked $K_s$ (red), $J$ (green) and $Y$ (blue) images. The image is $30\times30$ arcsec with North up and East to left.}
   \label{fig:rp}
\end{figure}

In the VMC colour-composite image PNe routinely appear as resolved red sources due to nebular emission lines of Br$\gamma$, He I 2.058 $\mu$m and 2.112 $\mu$m, and the H$_2$ molecular series (e.g. Hora et al. 1999). Figure \ref{fig:rp} shows RP1037 as a typical example of a resolved VMC PN detection. As part of an initial investigation into detecting new PNe like RP1037, we visually searched stacked VMC tiles as the VMC colour-composite in one \textsc{ds9} frame and a $K_s-Y$ image in another frame. The latter of which being sensitive to strong $K_s$ sources. During the search we came across an object in the 8\_8 tile with an unusual pink colour quite unlike others in the whole tile. As this colour is shared by some other bright PNe (e.g. SMP4, SMP6 and SMP30; see Miszalski et al. 2011), it was an excellent candidate for follow-up. The only previous reference to the object was made by Gruendl \& Chu (2009) who classified it as a possible AGB star.

\section{Results}
\label{sec:results}
\subsection{Four candidate PNe}
\begin{figure*}
   \begin{flushleft}
      \hspace{2cm} \includegraphics[scale=0.215]{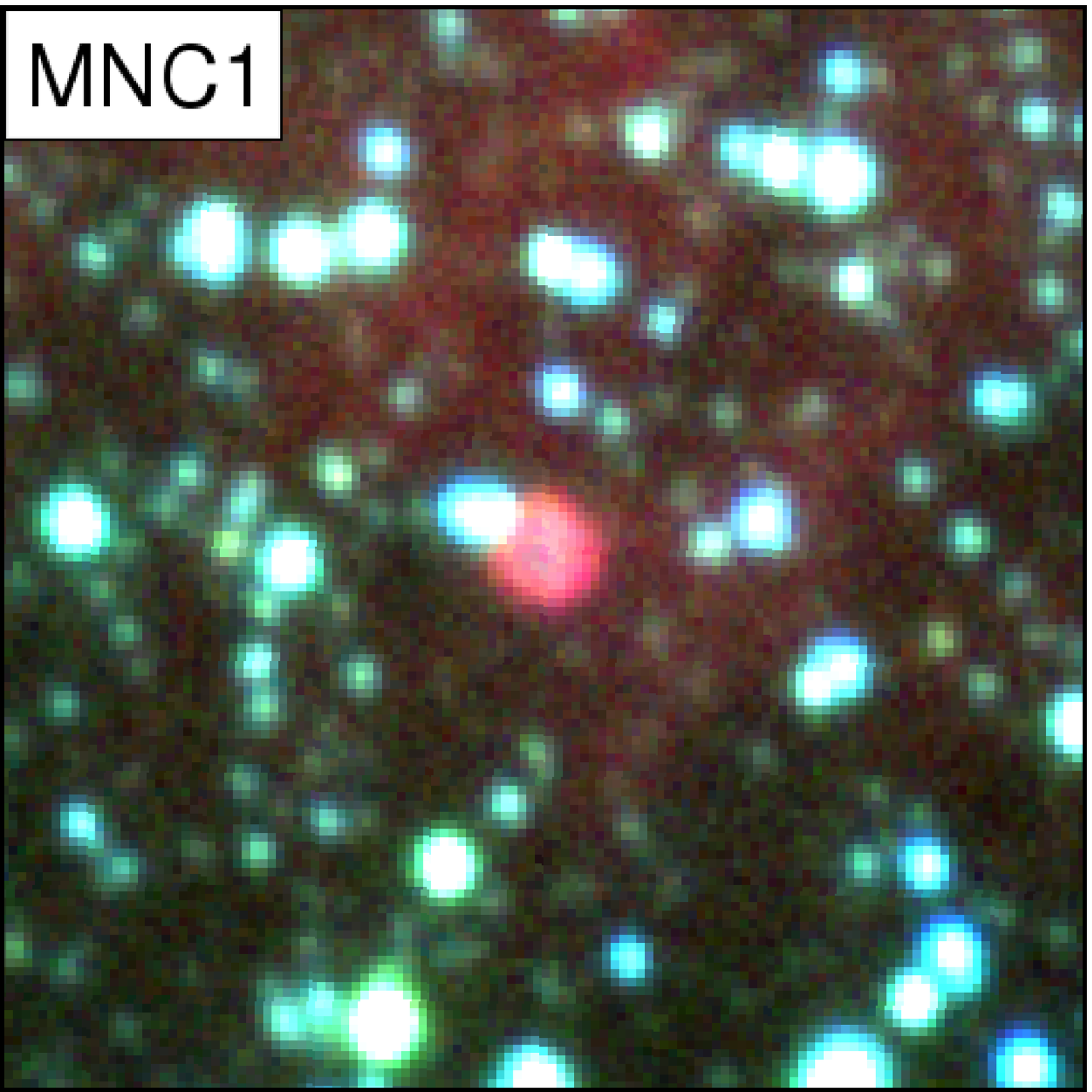}
   \includegraphics[scale=0.215]{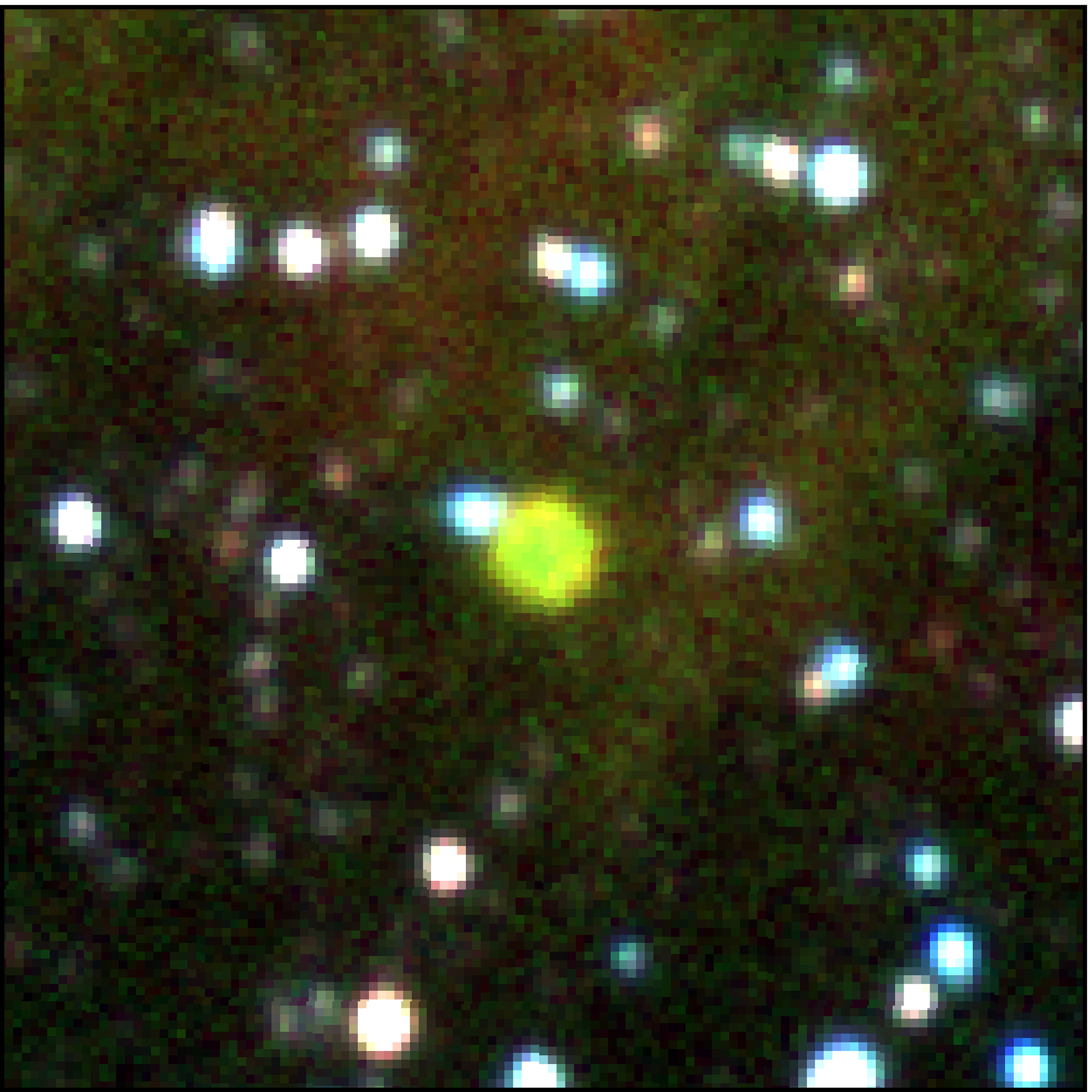}
   \includegraphics[scale=0.288]{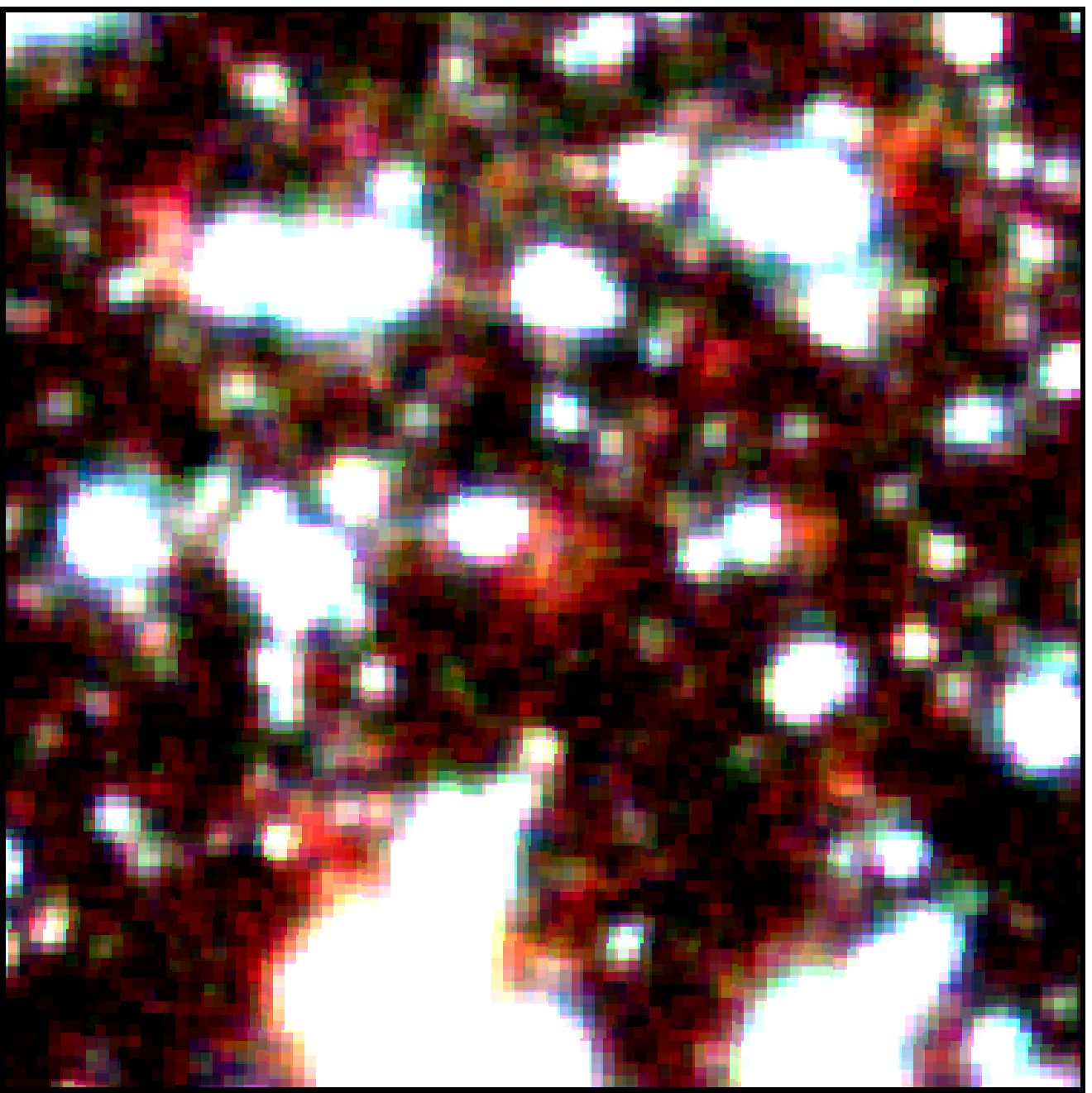}\\
\hspace{2cm} 
   \includegraphics[scale=0.215]{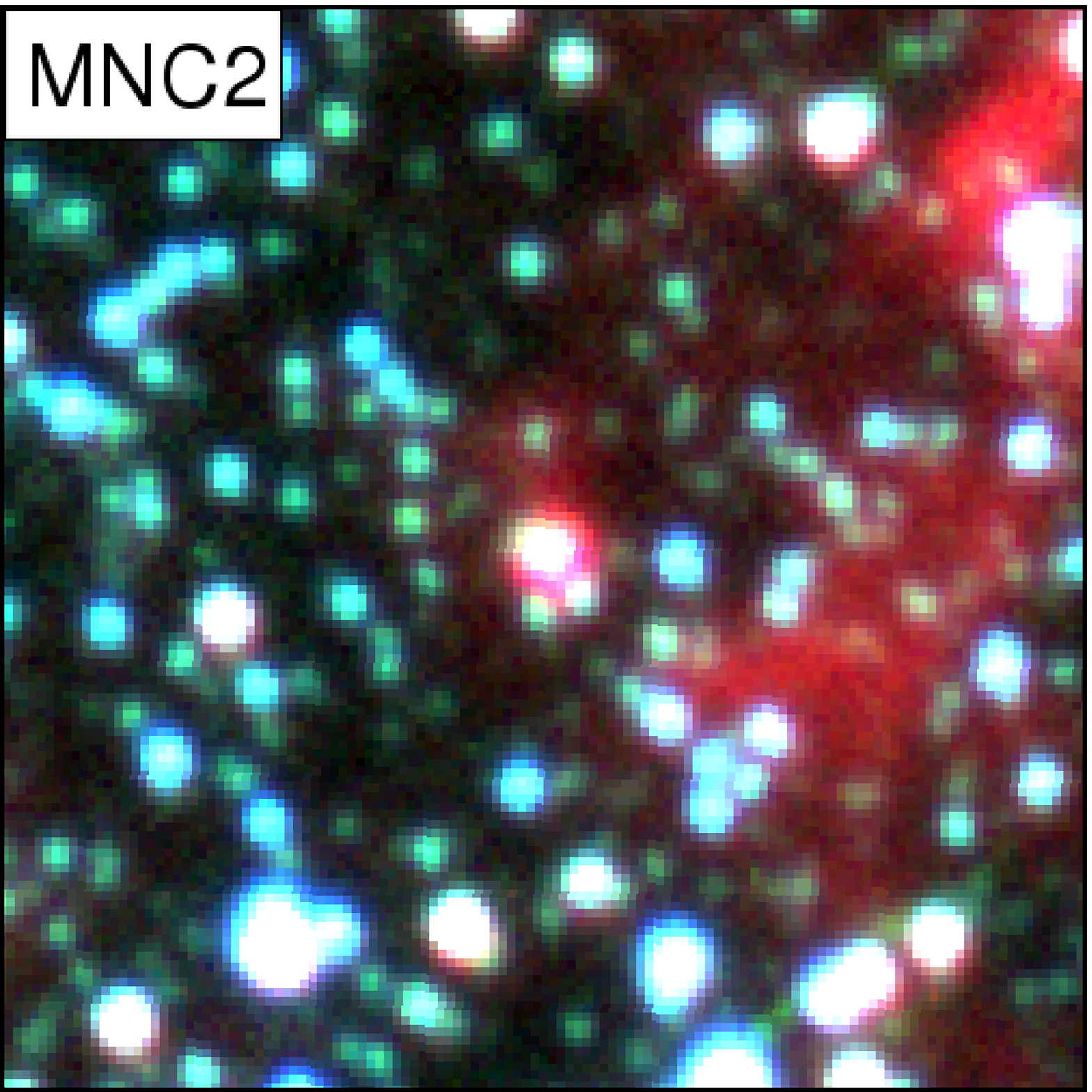}
   \includegraphics[scale=0.215]{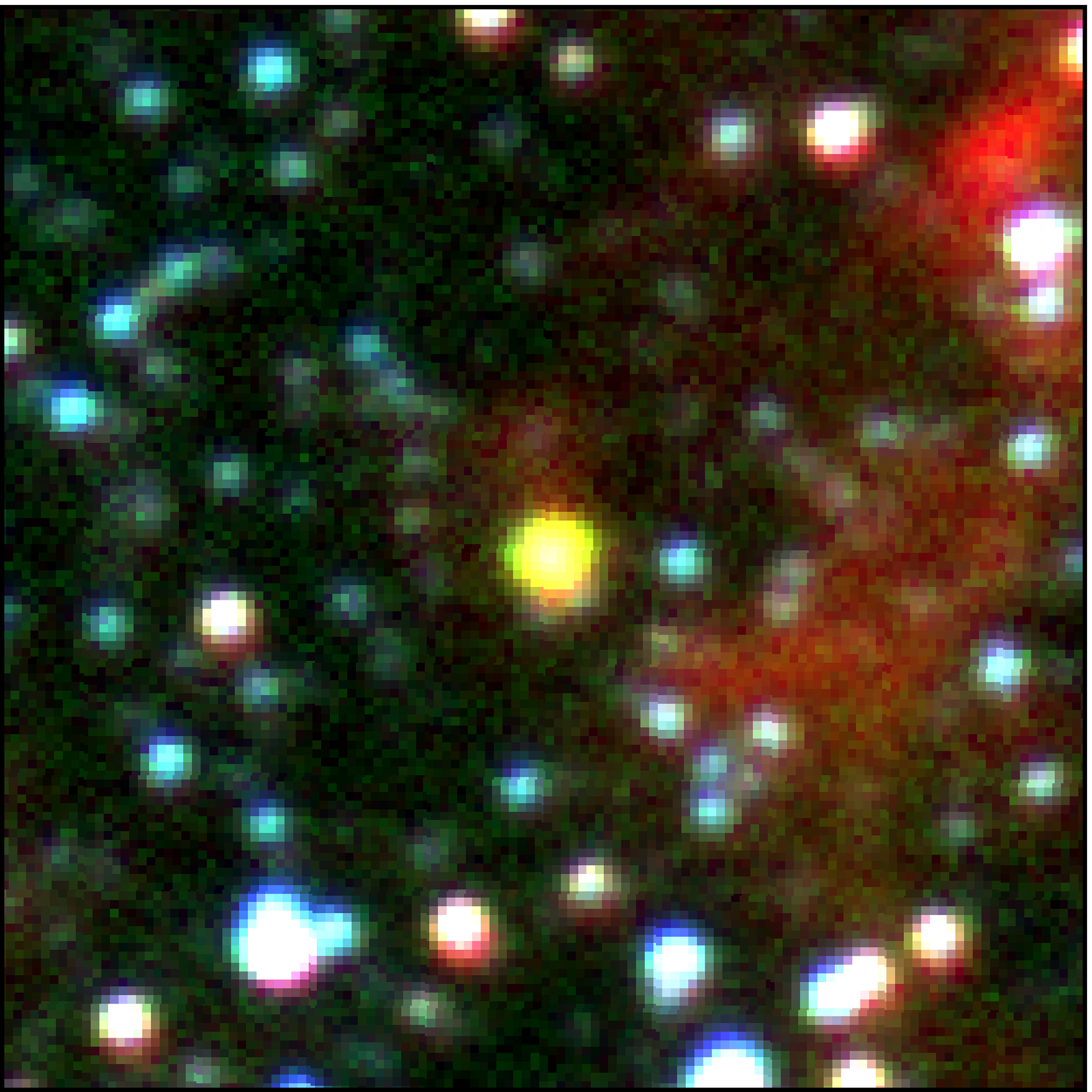}
   \includegraphics[scale=0.288]{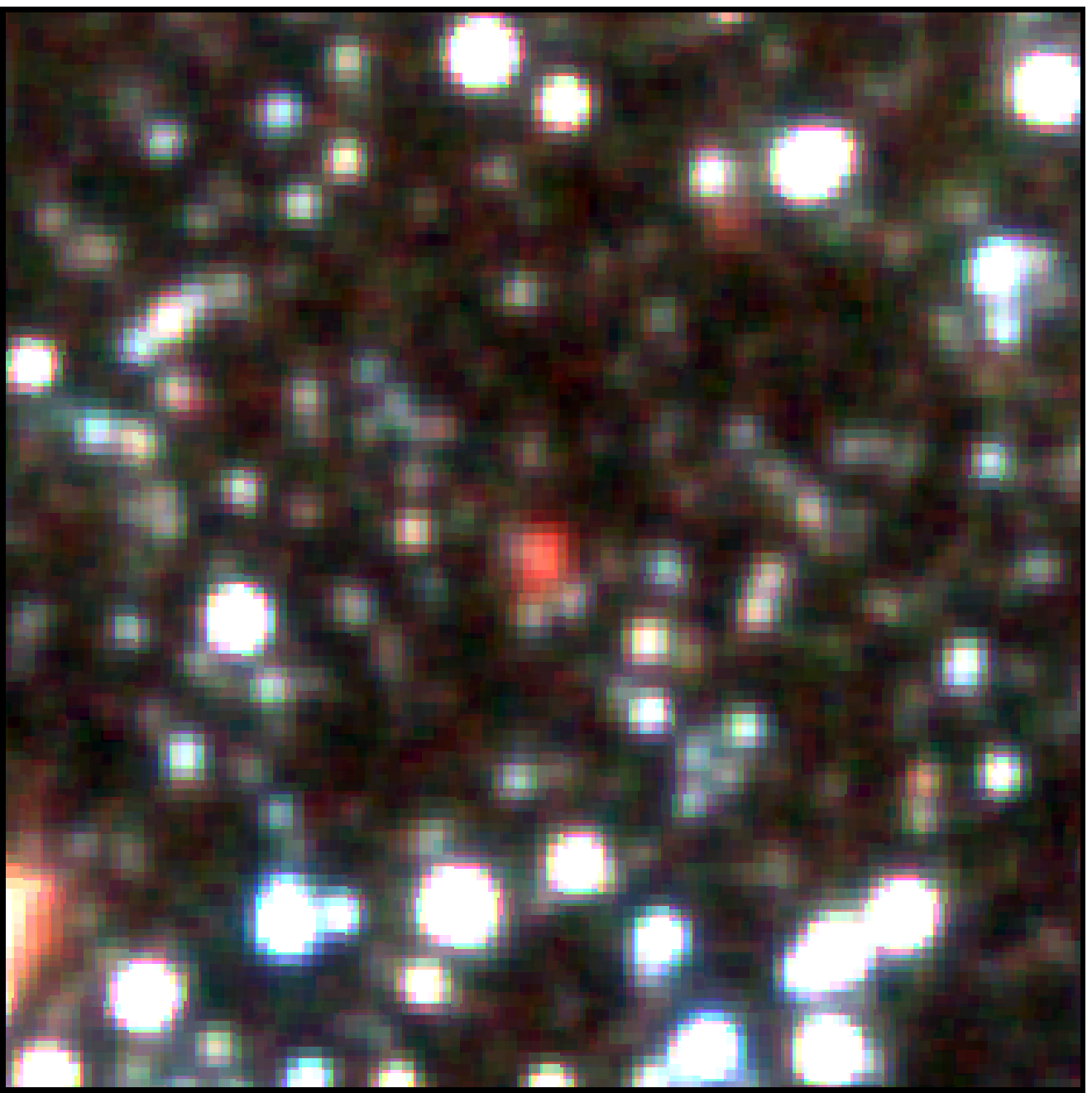}
   \includegraphics[scale=0.288]{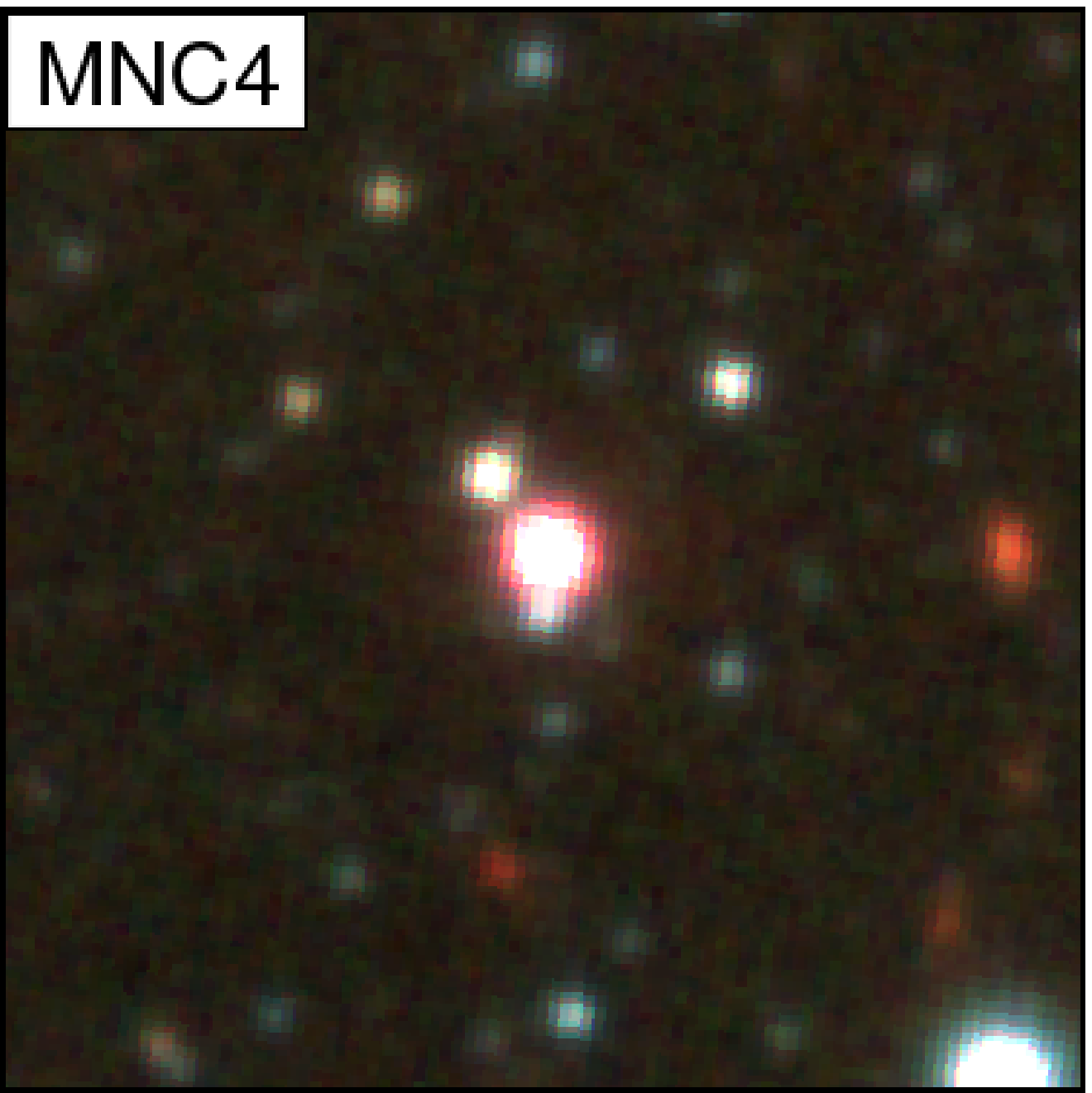}
   \\
 \hspace{2cm}   
   \includegraphics[scale=0.215]{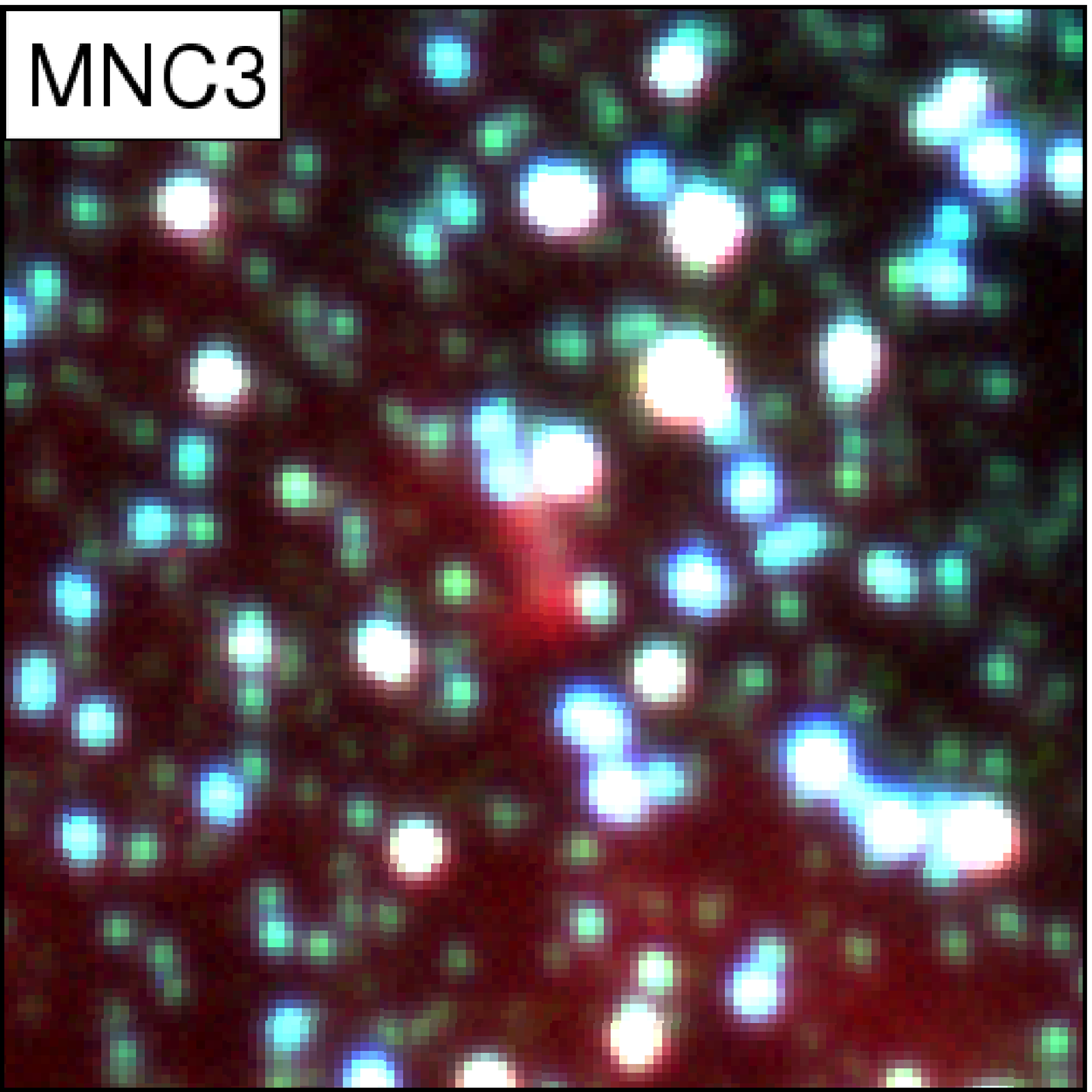}
   \includegraphics[scale=0.215]{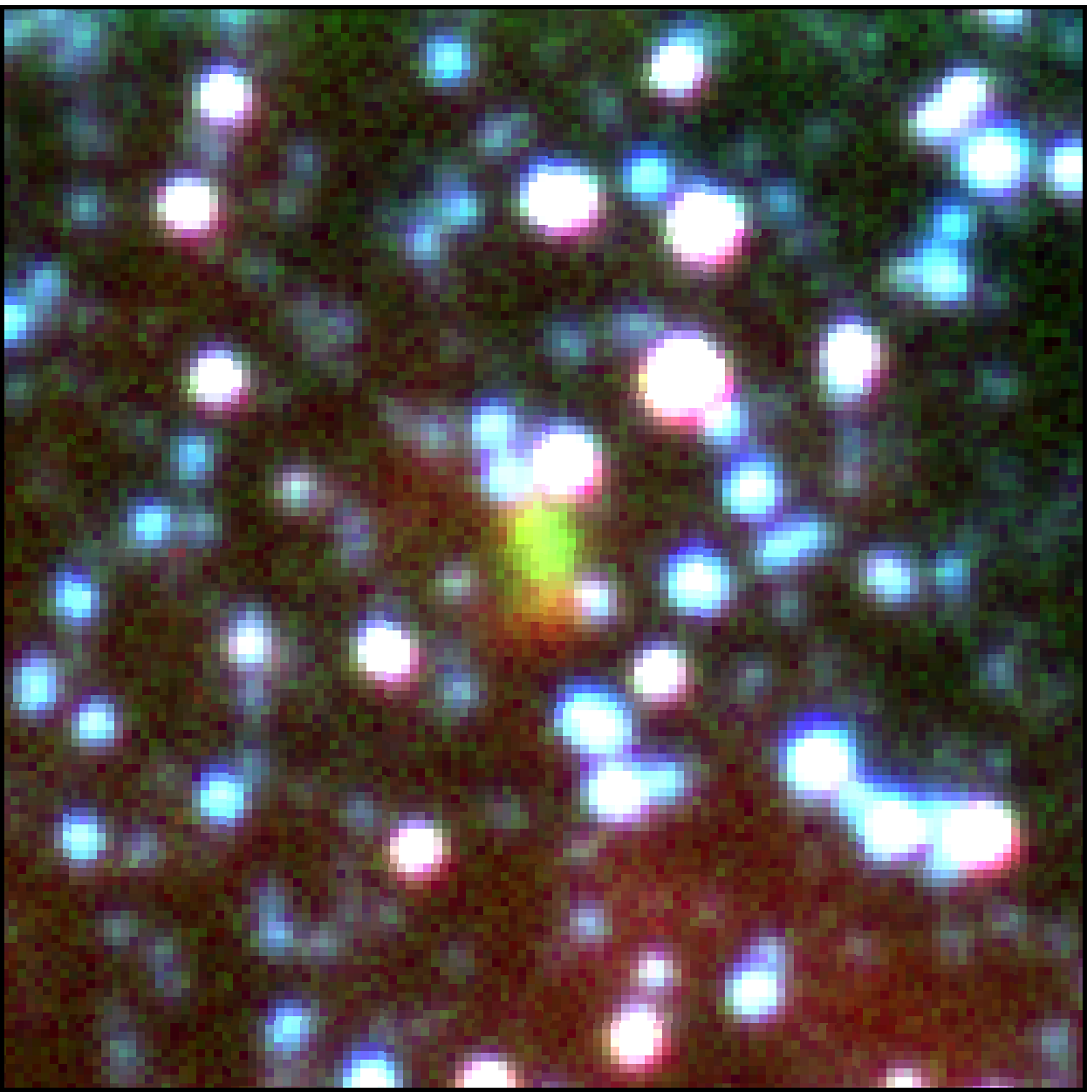}
   \includegraphics[scale=0.288]{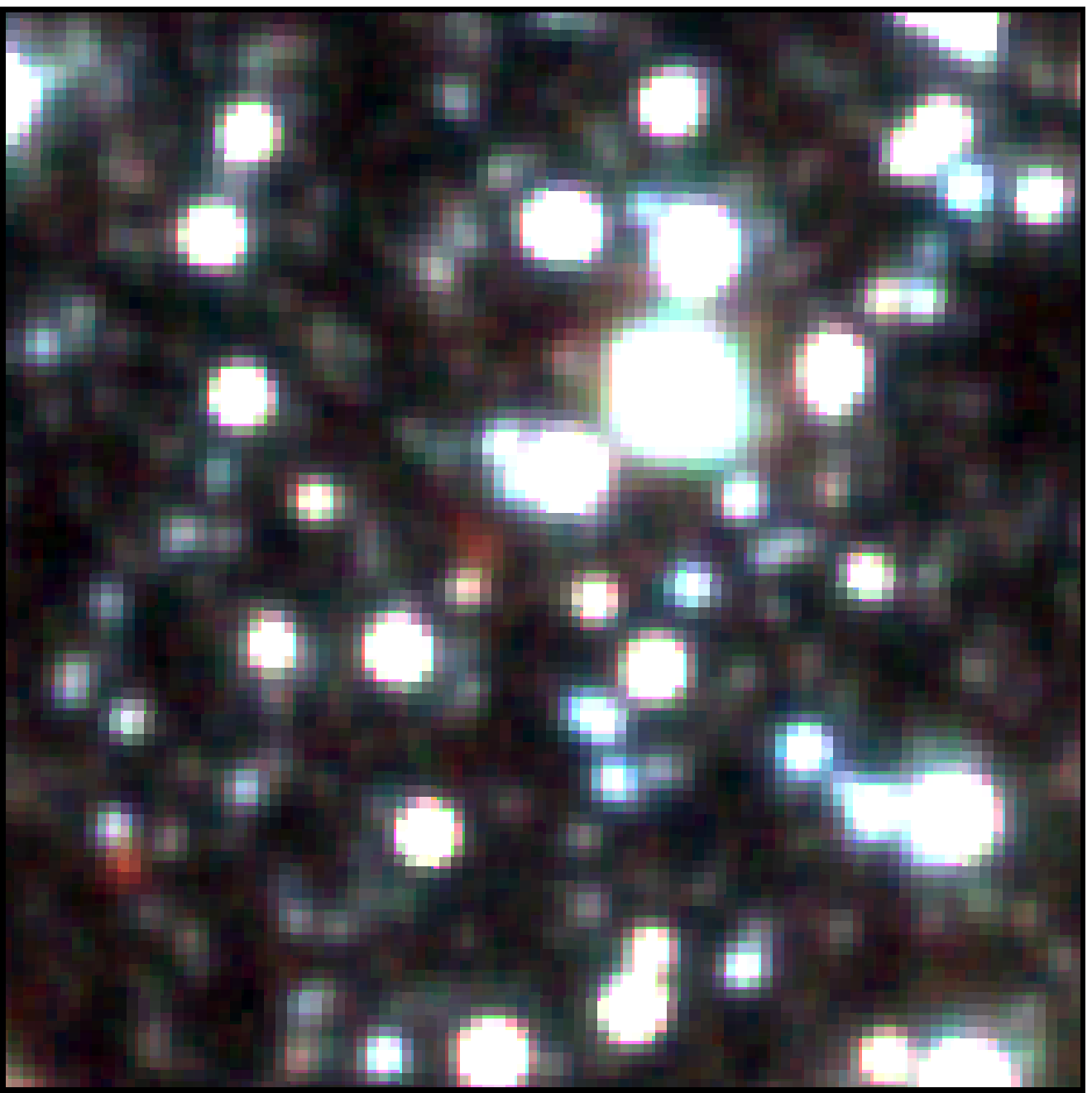}\\
   \end{flushleft}
   \caption{Optical and NIR colour-composite images of the four new PNe. (first column) H$\alpha$ (red), $V$ (green) and $B$ (blue). (second column) H$\alpha$ (red), [O~III] (green) and $B$ (blue). (third and fourth columns) VMC $K_s$ (red), $J$ (green) and $Y$ (blue). MNC4 only has VMC coverage and each image is $30\times30$ arcsec with North up and East to left.}
   \label{fig:new}
\end{figure*}

Four new PNe candidates were found whose basic properties are given in Tab. \ref{tab:new}. These properties are their names (after the authors Miszalski, Napiwotzki \& Cioni), parent WFI or VMC field, PN status (Sect. \ref{sec:spec}), equatorial coordinates, H$\alpha$ diameters, integrated [O~III] fluxes and magnitudes (see Sect. \ref{sec:fluxes}), heliocentric radial velocities (HRV, see Sect. \ref{sec:spec}) and morphologies. Figure \ref{fig:new} displays their WFI and VMC images. Two colour composites of H$\alpha$ (red), $V$ (green) and $B$ (blue), and H$\alpha$ (red), [O~III] (green) and $B$ (blue), were chosen to highlight the H$\alpha$ and [O~III] emission, respectively. The $K_s$ (red), $J$ (green) and $Y$ (blue) components of the VMC colour-composite are stacked images created by averaging individual images extracted from the reduced frames or `paw-prints' (Irwin et al. 2004; Cioni et al. 2011).

MNC1--3 are resolved in the WFI images and MNC1 and MNC2 in particular have strong [O~III] detections. In [O~III] and H$\alpha$ MNC1 seems to be round, but in $K_s$ it appears to have a bipolar internal structure (often called a `bipolar core') and higher spatial resolution imaging would be helpful to confirm this. MNC3 has a weaker [O~III] core embedded in an elongated H$\alpha$ nebulosity. Based on the available imaging it is difficult to distinguish between an enhancement in the diffuse HII background and an intrinsic PN since the ionisation structure resembles bipolar PNe (Corradi \& Schwarz 1995). MNC4 is unresolved in the VMC images. Except perhaps for MNC3, MNC1 and MNC2 are retrospectively detected in the RP2006a stack (Reid \& Parker, private communication), but were not selected as candidate PNe because of a combination of their faintness, small angular diameter and most importantly their complex H$\alpha$ background. If they were located in another region of the LMC besides 30 Doradus, then they would surely have appeared in RP2006b. MNC4 is located well outside the RP2006a survey footprint.

\begin{table*}
   \centering
   \caption{Basic properties of the four new PNe found.}
   \label{tab:new}
   \begin{tabular}{llllllllll}
      \hline\hline
      Name & Field & PN Status & RA & Dec. & Diameter & $\log F_{5007}$ & $m_{5007}$ & HRV & Morphology\\
      &       &  & (J2000)  & (J2000) & (\arcsec) & (erg s$^{-1}$ cm$^{-2}$) &  & (km s$^{-1}$) & \\
      \hline
      MNC1 & 30Dor1 & True & 05 42 46.80 & $-$69 20 30.0 & 3.4 & $-$13.60 & 20.26             &  $282$ & bipolar core\\      
      MNC2 & 30Dor3 & True & 05 42 43.97 & $-$69 35 34.0 & 2.5 & $-$13.57 & 20.18             &  $269$ & round/unresolved \\ 
      MNC3 & 30Dor3 & Possible & 05 41 28.35 & $-$69 43 53.0 & $1.9\times4.3$ & $-$13.85 & 20.88  &  $291$ & bipolar? \\         
      MNC4 & 8\_8 & True & 06 00 59.20 & $-$66 36 15.3 & - & - & -                            &  $269$ & unresolved\\        
      \hline
   \end{tabular}
\end{table*}

\subsection{WiFeS spectroscopy}
\label{sec:spec}
All our candidates were observed with the Wide Field Spectrograph (WiFeS; Dopita et al. 2007, 2010) on the Australian National University 2.3m telescope at Siding Spring Observatory. WiFeS is an integral field spectrograph with a $25\times38$\arcsec\ field-of-view that is sampled with 1\arcsec\ slitlets. We used the $B3000$ and $R3000$ volume-phase holographic gratings to achieve continuous spectral coverage from 3200--9800 \AA\ in a dual-beam configuration at a resolving power of 3000. Each object was observed for 10 minutes by one of us (JN) on 31 January 2011. At this time only half the detector space was operational and this resulted in 12 slitlet spectra. The WiFeS \textsc{IRAF} package (Dopita et al. 2010) was used to perform basic calibrations of the data before the 2D spectrum of each slitlet was cleaned of cosmic ray events. The highest signal-to-noise (S/N) 2D spectra centred on the objects were averaged from which sky-subtracted, one-dimensional spectra were extracted. No flux calibration was applied to the data. Radial velocities of our objects were also measured from the spectra using the \textsc{IRAF} package \textsc{emsao} (Kurtz \& Mink 1998). Table \ref{tab:new} lists the HRVs which are a weighted mean of blue and red measurements. The values are typical of LMC PNe (RP2006b) and we estimate their errors to be 5--10 km/s.

\begin{figure*}
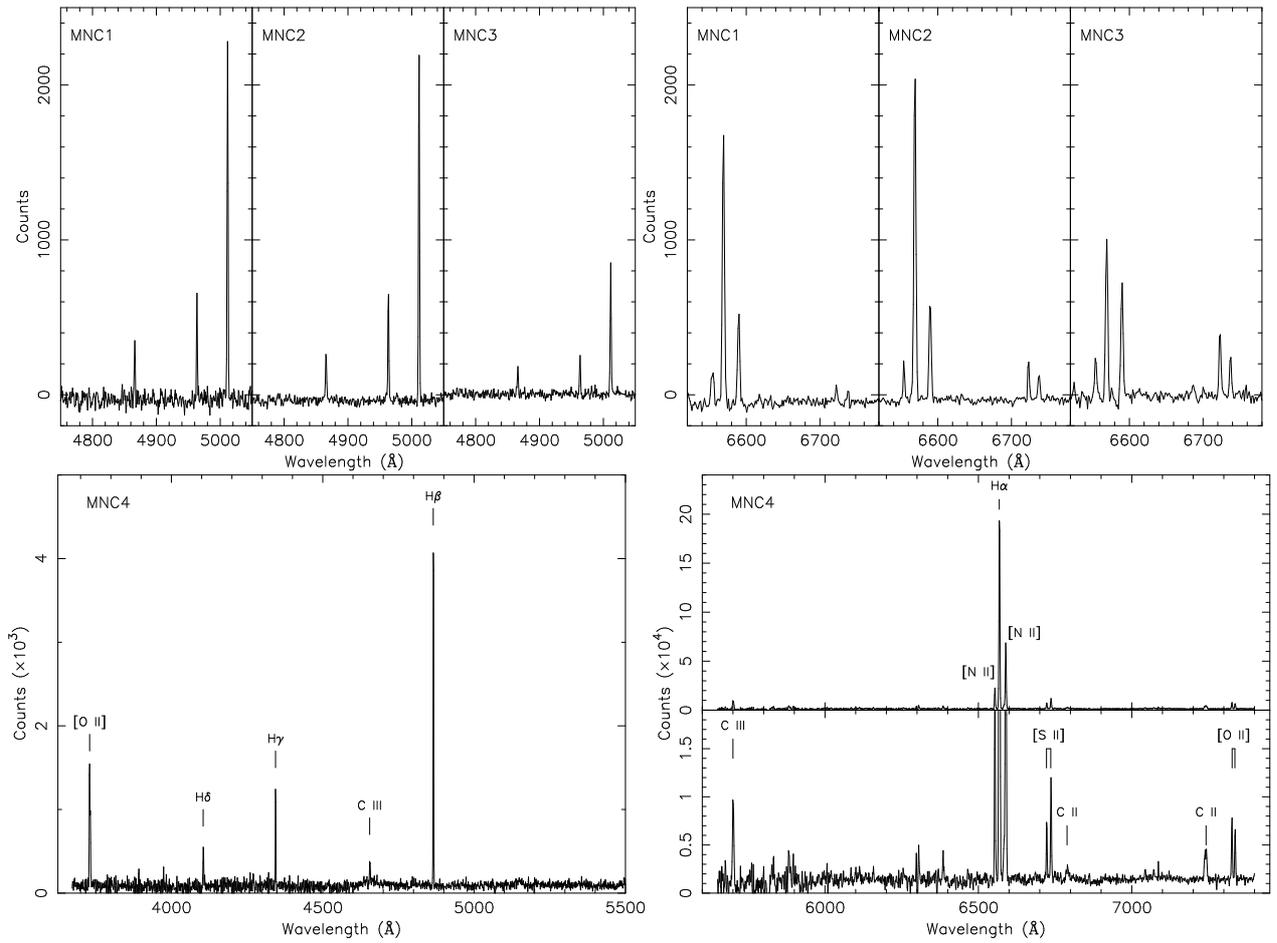

   \begin{center}
      \includegraphics[scale=0.35,angle=270]{mncB.ps}
      \includegraphics[scale=0.35,angle=270]{mncR.ps}
      \includegraphics[scale=0.35,angle=270]{mnc4B.ps}
      \includegraphics[scale=0.35,angle=270]{mnc4R.ps}
   \end{center}
   \caption{WiFeS spectroscopy of our sample.}
   \label{fig:spec}
\end{figure*}

Figure \ref{fig:spec} shows the reduced spectra of all four new PNe. The depth achieved in MNC1--3 is adequate to detect only the brightest emission lines and their spectra offer little more than confirmation of the ESO WFI imaging. In the case of MNC3 the emission line ratios of log(H$\alpha$/[N~II])=0.02 and log(H$\alpha$/[S~II])=0.17 are closer to those of supernova remnants or possibly diffuse HII regions than PNe (Frew \& Parker 2010). We can therefore only classify MNC3 as a possible PN until deeper longslit spectroscopy or higher resolution imaging can better trace the nature and spatial extent of the emission surrounding the compact [O~III] source.

MNC4 shows many more lines in its spectrum. From the central star we have identified stellar emission lines of C~II $\lambda$6783.9, $\lambda$7231.3, $\lambda$7236.4 and C~III $\lambda$4650.2 and $\lambda$5695.9. The strength of the C~III lines and the absence of C~IV emission lines suggests a [WC9]--[WC11] classification for the Wolf-Rayet central star (Acker \& Neiner 2003). A definitive classification cannot however be reached until deeper spectroscopy is obtained that reaches the continuum level. The discovery of a new [WC] central star in the LMC brings the total to only six known in the MCPNe population (Leisy \& Dennefeld 2006). Central stars with [WC9] or later classification are too cool to excite [O~III] emission in their surrounding nebulae, explaining its absence in MNC4. Once the central star evolves further it will heat up and ionise most or all of the now visible [O~II] emission into [O~III] to produce a more typical PN spectrum.

\subsection{[O~III] fluxes}
\label{sec:fluxes}
The integrated [O~III] flux is an important quantity for any extragalactic PN so that it may be added to the PNLF.
As the WFI data were taken in non-photometric conditions the measurement of [O~III] fluxes for the new PNe requires a calibration based on PNe with known fluxes. RP2010 published an extensive catalogue of suitable [O~III] fluxes for all PNe that overlap with the WFI observations. We adopt an error of 0.2 dex as recommended by RP2010 that includes all sources of error. 
The calibrators are spread across the four sub-fields as follows: MG68 (30Dor1), SMP78 and MG60 (30Dor2), MG73 (30Dor3) and RP1037, MG65, Sa122, SMP77 and RP789 (30Dor4). MG76 unfortunately fell into an inter-chip gap at the edge of the combined [O~III] image of the 30Dor1 sub-field. We performed aperture photometry on the [O~III] image of each PN using a circular aperture that included the most flux taking care as best we could to avoid field stars. Sky subtraction was performed by subtracting the average counts from a set of identical sized apertures. 

The measured counts may include nebular or stellar continuum contributions and this is especially the case for the brightest PNe. With only $B$ and $V$ images available these contributions cannot be safely subtracted as they both include substantial nebular contributions. Instead, we have removed the brightest two PNe SMP77 and SMP78 from our calibrator list and safely assume that the continuum contributions for MG60, MG65 and Sa122 are smaller than the errors (0.2 dex). We also excluded MG68 which was found to not fit the trend (see below).

Figure \ref{fig:fit} shows the derived calibration after application of a linear least squares fit. The well-behaved trend is a testament to the careful [O~III] flux measurements performed by RP2010. The fluxes in Tab. \ref{tab:new} were derived from the fit after measuring their counts and have an associated uncertainty of 0.07 dex from the standard error of the fit. This value is also twice the measured $\sigma$ from the fit-subtracted residuals. Jacoby (1989) $m_{5007}$ magnitudes suitable for the PNLF were also calculated as usual via the relation $m_{5007}=-2.5 \log F_{5007} - 13.74$ and placed in Tab. \ref{tab:new}. 

MG68 was excluded from the fit because its published flux falls 2$\sigma$ (0.4 dex) lower than the expected value of log F$_{5007}=-12.74\pm0.07$ dex. This may be explained by either (a) the non-photometric conditions of the data, or (b) a real underestimation by RP2010. It does not seem possible to distinguish between the two as there are no other calibrator PNe besides MG68 in the 30Dor1 sub-field. If (b) is true, then this could be explained by the strong spatial variation of [O~III] in MG68 which is not accommodated by the small 2\arcsec\ fibre aperture of RP2010 (see Miszalski et al. 2011).

\begin{figure}
   \begin{center}
      \includegraphics[scale=0.37,angle=270]{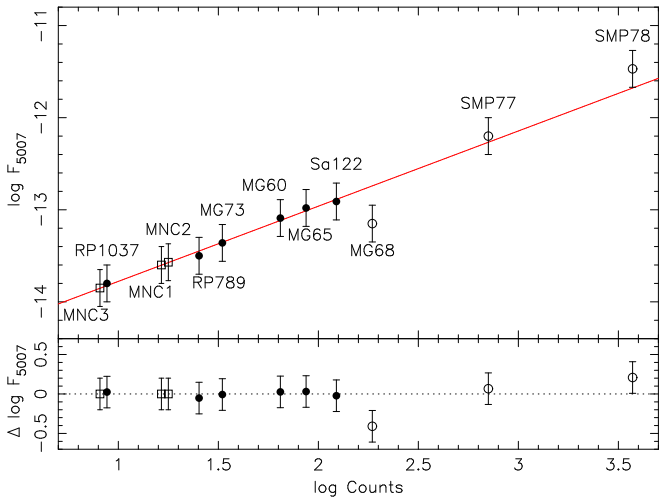}
   \end{center}
   \caption{Calibration relation for the WFI [O~III] data based on calibrator PNe (filled points). The three new PNe are shown by squares and circles show PNe excluded from the fit.}
   \label{fig:fit}
\end{figure}

\section{Conclusions}
\label{sec:end}
A systematic search for new PNe was conducted in a $63\times63$ arcmin$^2$ region near the star forming complex 30 Doradus using publicly available deep ESO WFI $B$, $V$, [O~III] and H$\alpha$ imaging. Two bona-fide PNe (MNC1 and MNC2) and one possible PN (MNC3) were identified and their basic properties presented. The [O~III] images were calibrated in flux using previously known PNe with literature fluxes. At 6--7 magnitudes fainter than the PNLF cut-off, the new discoveries indicate the PN population in the 30 Doradus region is relatively complete to this depth. There may however be some PNe totally embedded in the nebula emission and velocity-sensitive techniques such as cross-dispersed or fabry-perot imaging may be required to detect them (Jacoby \& De Marco 2002). 
The main result of our survey is that there remains a small population of faint PNe of small angular size awaiting discovery in the LMC. At 2--3 PN per square degree, with identical observations of the inner $5\times5$ degrees coinciding with the RP2006a survey footprint, we would expect to reveal 50--75 new PNe 6--7 mag fainter than the bright-end of the PNLF. This is a substantial population that should be factored into estimates of the total LMC PN population. 

   An additional bona-fide PN, MNC4, was found serendipitously from visual examination of the deep 8\_8 tile from the near-infrared VMC survey. Spectroscopic observations show the object to be a PN with a rare [WC9]--[WC11] Wolf-Rayet central star. There are now at least six [WR] central stars in MCPNe (Leisy \& Dennefeld 2006). This object is located outside the RP2006b survey footprint and clearly demonstrates that the VMC survey can find new PNe that are underrepresented in the current population. 
   
   Estimating the theoretical, or indeed, observed LMC PN population is at present highly uncertain. The theoretical estimate of $1000\pm250$ PNe (Jacoby 1980; Peimbert 1990) contains a large uncertainty that could easily be higher. Emphasis must therefore instead be placed on producing a robust tally of observed PNe. Our survey of a small region is a positive step in this direction, but the VMC survey will have a much larger impact on the tally. We have demonstrated here with MNC4 that it is possible to find new PNe from the VMC survey alone. The VMC survey is also well-suited to the removal of contaminating objects that fall into H$\alpha$ selected samples such as RP2006b (e.g. Frew \& Parker 2010). Initial results indicate a very high contamination fraction of $\sim$50\% for 98 previously catalogued PNe that are mostly located near 30 Doradus (Cioni et al. 2011; Miszalski et al. 2011). It is plausible that up to 170 PNe classified as `likely' and `possible' by RP2006b could eventually be removed with the assistance of further VMC observations. This would drop the PN population down to $\sim$620 PNe which includes our potential survey contribution of 50--75 PNe. 

\begin{acknowledgements}
   The authors wish to thank George Jacoby for very helpful discussions regarding the PNLF, Peter Wood for coordinating the WiFeS spectroscopic follow-up and Catherine Farage for her exceptional help that allowed for the WiFeS data to be reduced. We also thank Warren Reid and Quentin Parker for sharing their survey data of our new discoveries and the anonymous referee whose comments helped improve this paper. This research has made use of SAOImage DS9, developed by Smithsonian Astrophysical Observatory.
\end{acknowledgements}

\end{document}